# ABSTRACT

Designing and developing Artificial Intelligence controllers on separately dedicated chips have many advantages. This report reviews the development of a real-time fuzzy logic controller for optimizing locomotion control of a two-wheeled differential drive platform using an Arduino Uno board. Based on the Raspberry Pi board, fuzzy sets are used to optimize color recognition, enabling the color sensor to correctly recognize color at long distances, across a wide range of light intensity, and with high fault tolerance.

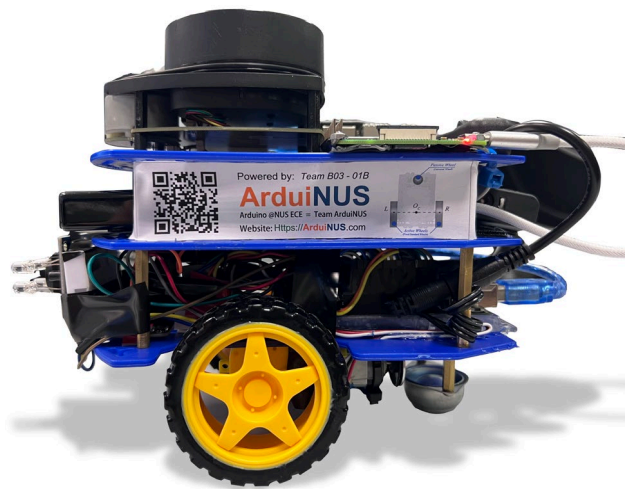

This is a Side View of ALEX

Designed and assembled by **Team ArduiNUS**

Due to the superiority of our ALEX in algorithm and hardware design,
this robot is preserved in the Digital Systems & Applications laboratory for future engineers to study.

# INTRODUCTION

This First Year Project (FYP) Report encapsulates my comprehensive academic journey throughout my first year at the National University of Singapore, aiming to illustrate a careful integration of concepts from modules ME3243/EE3305 Robotics System Design, EE4305 Fuzzy/Neural Systems for Intelligent Robotics, and CG2111A Engineering Principles and Practice. Through the practical application of these modules on the Arduino Uno & Raspberry Pi (RPi) robotic platform (Referred to as Robot ALEX in module CG2111A), I have engineered a rescue robot governed by fuzzy control algorithms.

In Module CG2111A, the rescue site is conceptualized as a Maze, where both the rescue robot and human operator do not have prior knowledge of the Maze's terrain. They can only identify the terrain through an onboard LiDAR system. The victim is represented by a green or red cylindrical object that can appear anywhere in the Maze. Since there are cylinders of other colors present in the Maze, the robot needs to determine the color of the cylinders to confirm whether they represent victims.

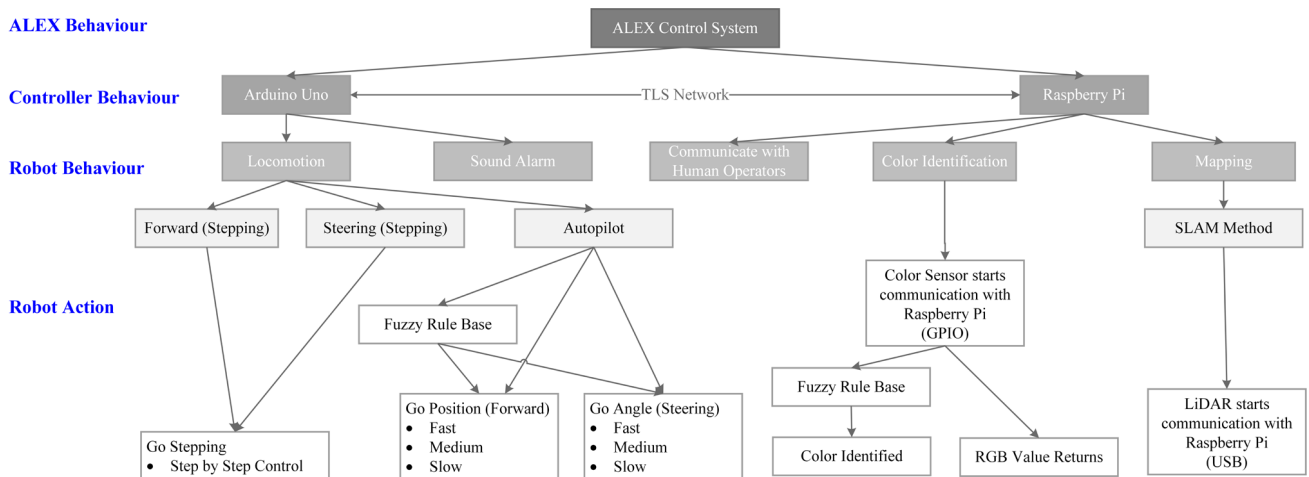

**Fig 0.1** Behavior architecture for ALEX

The fuzzy control algorithm optimizes the robot's locomotion and ability to identify victims. Chapters 1 - 3 focus on locomotion, and Chapter 4 analyzes color recognition. Chapter 1 uses kinematic models and constraints to analyze the locomotion of the two-wheel differential drive platform, represented by the ALEX robot configuration. Chapter 2 employs the Proportional–Integral–Derivative (PID) control algorithm by using the equation of motion derived in Chapter 1 to optimize the ALEX robot's movement. Chapter 3 implements fuzzy control algorithms to refine the control accuracy of the PID algorithm. When the robot approaches a victim, it will utilize an optical color sensor to assess the victim's condition. Chapter 4 illustrates the optimization of the color-recognition process using a fuzzy algorithm.

The acknowledgment, *Guoyi's First Adventure in Robotics*, is the final part of this paper, it will narrate my learning journey in my Year 1 at the National University of Singapore.

# INDEX

# Chapter 1 Analysis of Locomotion

As shown in Figure 1.1, the ALEX robot controls its locomotion by adjusting the driving velocity of two active wheels on both sides: If the left and right active wheels have the same driving velocity, then the robot stays static or travels in rectilinear motion; if the left and right active wheels have different driving velocities (namely, differential velocity), then the robot performs circular motion. ALEX is a two-wheeled differential robot based on the abovementioned motion characteristics. This research only examines the motion control system of two-wheeled differential robots represented by the ALEX configuration.

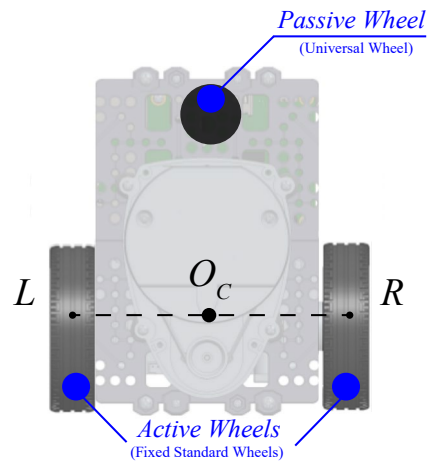

**Fig 1.1** Structural diagram is the top view of ALEX. This configuration consists of a left active wheel without steering capacity, denoted by $L$, a right active wheel $R$, and a passive wheel. The $O_C$ point is the robot's spin center point, which is always located at the center point of the line connecting the pivot points of the $L$ and $R$ wheels.

Before applying the fuzzy-PID theory to control ALEX, we first need to analyze its equation of motion to design its motion control system. The equation of motion consists of the forward equation of motion and the inverse equation of motion. The forward equation of motion calculates the movement velocity of the robot's spin center point using known driving velocities of the left and right active wheels. In contrast, the inverse equation of motion calculates the driving velocity of the left and right active wheels using the known movement velocity of the robot's spin center point. This chapter will analyze the solutions to the forward and inverse equation of motion of ALEX. In Chapter 2, the PID control theory and the inverse equation of motion are applied to design the robot's motion control system. Chapter 3 applies the forward equation of motion and the fuzzy control theory in tuning the motion control system.

Prior to solving the equation of motion, we need first establish the Global Reference Frame $\xi_I$ and the Robot Local Reference Frame $\xi_R$, as shown in Figure 1.2, to set the parameters of these equations.

## 1.1 Establishing the Global Reference Frame

In the Global Reference Frame, robotic motion can be regarded as a rigid body's motion on a two-dimensional plane, which can be decomposed into a translational movement with two degrees of freedom (*DOF*s) and a rotary motion with one *DOF*. Therefore, the robotic Global Reference Frame has three theoretical degrees of freedom ($DOF_{workspace}$). The robot's locational parameters in the Global Reference Frame can be set as a vector $\xi_I = [x_I \quad y_I \quad \theta_I]^T$ whose dimensions are the same as its degrees of freedom. Then, the velocity parameter of the robot in the Global Reference Frame can be set as the vector $\dot{\xi}_I = [\dot{x}_I \quad \dot{y}_I \quad \dot{\theta}_I]^T$.

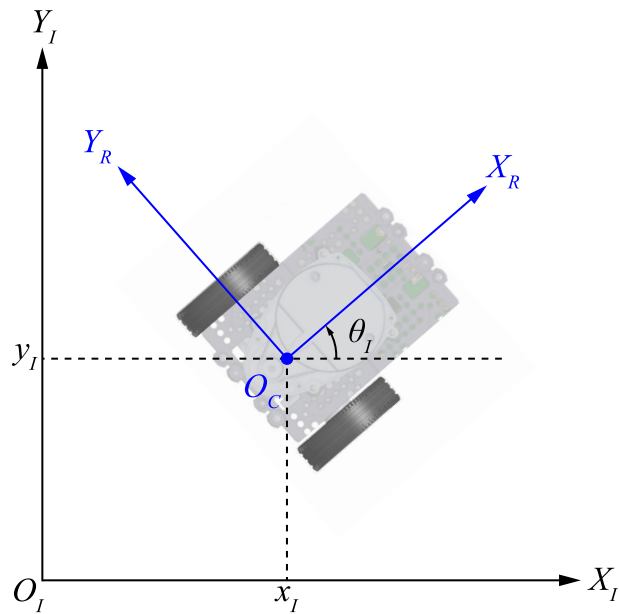

**Fig 1.2** Global Reference Frame $\xi_I$ and Robot Local Reference Frame $\xi_R$. Angle $\theta_I(\theta_I \in \{0 \leq \theta_I \leq 180°\})$ is the angle between the $X_I$ axis of the $\xi_I$ frame and the $X_R$ axis of the $\xi_R$ frame.

## 1.2 Establishing the Robot Local Reference Frame

The Robot Local Reference Frame $\xi_R$ is established by taking the spin center point of the robot as the origin. To satisfy the right-hand rule, the forward motion direction of the robot is defined as the positive direction of the $X_R$ axis, what is perpendicular to the left is the positive direction of the $Y_R$ axis, and the $Z_R$ axis is perpendicular to the $O - XY$ plane to the outward direction. To sum up, the velocity parameter $\dot{\xi}_R$ of the robot's spin center in the Global Reference Frame $\xi_I$ can be obtained from point multiplication of the velocity vector of the Global Reference Frame $\dot{\xi}_I$ and the orthogonal rotation matrix $^I_R R(\theta)$:

$$\dot{\xi}_R = {^I_R}R(\theta)\dot{\xi}_I = \begin{bmatrix} \cos\theta & \sin\theta & 0 \\ -\sin\theta & \cos\theta & 0 \\ 0 & 0 & 1 \end{bmatrix} \begin{bmatrix} \dot{x}_I \\ \dot{y}_I \\ \dot{\theta}_I \end{bmatrix} \quad (1)$$

By now, we have already obtained the equation of motion shown in Equation (1). However, such an equation of motion contains parameters of the Global Reference Frame that the robot cannot directly obtain. Therefore, we are unable to design the robot's motion control system using Equation (1). In reality, ALEX can only control the driving velocities of left and right active wheels, namely, $v_l$ and $v_r$. In addition, as the active wheels of ALEX are unable to slide along the direction perpendicular to their running direction, the robot is incapable of immediately change its motion along certain directions by altering the driving velocities of active wheels. This means the robot's *DOF* in motion is restricted by its mechanical structure. Therefore, to determine the dimensions of the equation of motion of the robot, we need to explore the actual *DOF* of the moving robot equivalent to the dimensions of the equation of motion (namely, the differential degree of freedom, *dDOF*).

## 1.3 Analysis of the Dimensions of Equation of Motion

To sum up, the robot's *dDOF* is not equal to its *DOF*<sub>workspace</sub> due to the constraint over the sliding of active wheels. The following equation can express the relationship between the constraint and *DOF*:

$$DOF_{workspace} = dDOF + C_f \tag{2}$$

Where $C_f$ is the sliding constraint degree, denoting how many directions along which the active wheels are subject to sliding constraint. The active wheel coordinate frame for ALEX $\xi_W$ is constructed, as shown in Figure 1.3. As can be known, the distance between the rotatory center point of active wheel $O_W$ and spin center point $O_C$ is constantly $l$. The angle between the robot body and wheel running direction is constantly $\alpha + \beta$. Thus, the linear velocity of active wheel $v_W$ can be expressed by:

$$v_W = \begin{bmatrix} \sin(\alpha + \beta) \\ -\cos(\alpha + \beta) \\ -l\cos(\beta) \end{bmatrix}^T {}_R^I R(\theta) \dot{\xi}_I = r_W \cdot \omega_W \tag{3}$$

Based on the constraints mentioned in Chapter 1.2, the normal velocity of $v_w$ can be expressed by:

$$\perp v_W = \begin{bmatrix} \cos(\alpha + \beta) \\ \sin(\alpha + \beta) \\ l\sin(\beta) \end{bmatrix}^T {}_I^R R(\theta) \dot{\xi}_I = 0 \tag{4}$$

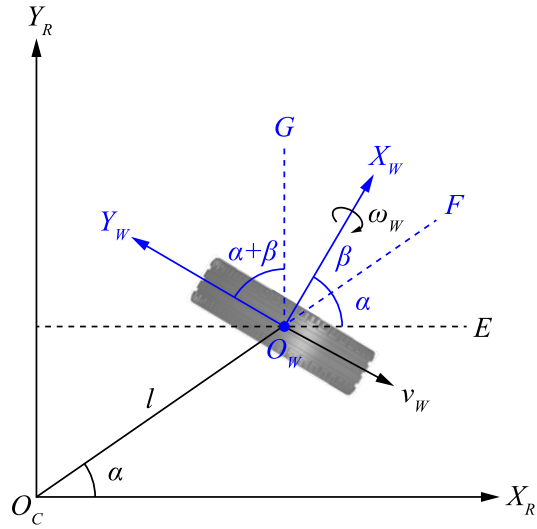

**Fig 1.3** The Robot Local Reference Frame $\xi_R$ and active wheel coordinate frame $\xi_W$. Line $E$ is set to cross the rotatory center point of active wheel $O_W$ and is parallel to the $X_R$ frame of the $\xi_R$ frame. The length of the segment $O_C O_W$ is set as $l$, and the radius of the active wheel is set as $r_W$. Half-line $O_C F$ is set as the extended line of $O_W O_C$, and half-line $O_W G$ is perpendicular to the line $E$. It is set that $\angle E O_W F = \angle O_W O_C X_R = \alpha$, $\angle X_W O_W F = \beta$.

As the sliding constraint capacity of the active wheel during motion only depends on wheel structure and the position where it is mounted on the robotic body, this parameter is always constant. To calculate the sliding constraint capacity of the active wheel more conveniently, we stack the reference frames of $\xi_I$ and $\xi_R$ to make their three axial directions align and share the same origin. At this point, the left wheel parameter is: $\alpha_l = -\frac{\pi}{2}, \beta_l = \pi, l_l = l$ and the right wheel parameters: $\alpha_r = \frac{\pi}{2}, \beta_r = 0, l_r = l$, then we have:

$$\perp v_w = \begin{bmatrix} 0 & 1 & 0 \\ 0 & 1 & 0 \end{bmatrix} \begin{bmatrix} 1 & 0 & 0 \\ 0 & 1 & 0 \\ 0 & 0 & 1 \end{bmatrix} \begin{bmatrix} \dot{x}_I \\ \dot{y}_I \\ \dot{\theta}_I \end{bmatrix} = \begin{bmatrix} 0 \\ 0 \end{bmatrix} \tag{5}$$

So, we can get $\dot{y}_I = 0$. This means the robot body cannot move along the direction perpendicular to the rolling direction of active wheel, and the ALEX structured robot has one degree of sliding constraint. According to Equation (2), we get:

$$dDOF = DOF_{\text{workspace}} - C_f = 3 - 1 = 2$$

To sum up, the actual DOF can be determined as two by analyzing the constraint and DOF of ALEX. Hence, the equation of motion of the ALEX structured robot can be expressed as a two-dimensional vector.

## 1.4 Analytical Equation of Motion

The two-wheeled differential model of ALEX is shown in Figure 1.4, where the instantaneous center of rotation (*ICR*) is the steering center point, denoting that the robot only performs rotary motion around the point. When velocities of left and right active wheels on the same rigid body are differential, this leads to the rotation of the rigid body in motion. At this point, if the steering radius $r_C$ is not zero, then the position of the rigid body is also changing and thus there is also translational movement. Given the sliding constraint on the robot body, the translational movement of the robot can only be fulfilled along the rolling direction of the wheel (the $X_R$ axis direction) instead of the $Y_R$ axis direction.

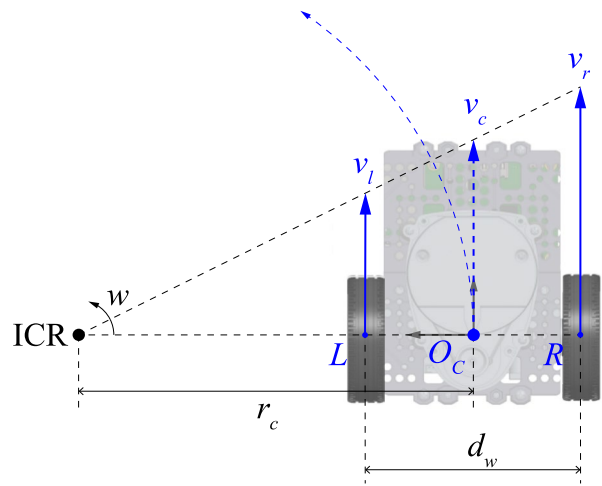

**Fig 1.4** ALEX two-wheeled differential model. The distance between *ICR* and $O_C$ is set as the steering radius $r_C$; $L$ and $R$ are set as the left and right active wheels of the robot, respectively; and the distance between $L$ and $R$ is $d_W$. $V_c$ represents the speed of the robot's center of rotation, while $V_l$ and $V_r$ denote the linear velocities of the left and right drive wheels, respectively.

Therefore, all points in the Robot Local Reference Frame $\xi_R$ can be described by the angular velocity and linear velocity along the $X_R$ axis. Based on parameters shown in Figure 1.4 and sliding constraint of ALEX derived in Chapter 1.3, the spin center point velocity can be described by the equation $\dot{\xi}_R = [v_c \quad w]^T$. As the linear velocity directions of the left and right active wheels are the same as the $X_R$ axis and the linear velocity direction is perpendicular to steering radius, the *ICR* is constantly located on the connecting line between points $L$ and $R$, and the specific location of *ICR* on line $LR$ is determined by the left and right active wheel velocity $[v_l \quad v_r]$. With $v = w \cdot r$, when $w$ is constant, $v$ is proportional to $r$. Hence, the velocities of points $L$, $R$ and $O_C$ can be expressed as:

$$w = \frac{v_c}{r_c} = \frac{v_r}{r_c + d_w/2} = \frac{v_l}{r_c - d_w/2} \qquad (6)$$

Based on the formula deformation of Equation (6), the angular velocity $w$ can be expressed as:

$$w = (v_r - v_l) / d_w \tag{7}$$

Further, the relationship between the linear velocity $v_c$ of the spin center point and the left and right active wheel velocity $[v_l \quad v_r]$ can be calculated by Equation (7).

$$v_c = (v_l + v_r) / 2 \tag{8}$$

## 1.5 Motion Equations and their Application

The *Forward Motion Equation* calculates the velocity of the spin center point $O_C$ based on the velocities of the left and right driving wheels. Combining equations (7) and (8), the positive motion equation can be expressed as:

$$\begin{bmatrix} v_c \\ w \end{bmatrix} = \begin{bmatrix} \dfrac{v_r + v_l}{2} \\ \dfrac{v_r - v_l}{d_w} \end{bmatrix} = \begin{bmatrix} 1/2 & 1/2 \\ 1/d_w & -1/d_w \end{bmatrix} \begin{bmatrix} v_r \\ v_l \end{bmatrix} \tag{9}$$

In real-world application, encoder will be mounted on both driving wheels of the differential robot, and the linear velocities $[v_l \quad v_r]$ of the two driving wheels can be calculated. At this point, the velocity of robot $O_C$ can be calculated by Equation (9).

The *Inverse Motion Equation* decomposes the velocity of spin center point $O_C$ into left and right driving wheel velocities, which can be expressed as:

$$\begin{bmatrix} v_r \\ v_l \end{bmatrix} = \begin{bmatrix} v_c + \dfrac{d_w}{2} w \\ v_c - \dfrac{d_w}{2} w \end{bmatrix} = \begin{bmatrix} 1 & d_w/2 \\ 1 & -d_w/2 \end{bmatrix} \begin{bmatrix} v_c \\ w \end{bmatrix} \tag{10}$$

The inverse motion equation is used to control robot $O_C$ to operate at the expected velocity $[v_c \quad w]^T$, that is, the theoretical velocities of the two driving wheels are calculated by Equation (10). In next chapter we will utilize the wheel encoder and inverse motion equation obtained in this chapter to optimize the robot's motion along the expected route in combination with the PID control theory.

# Chapter 2 Implementation of PID Control Algorithm

The drive platform of the ALEX robot incorporates two DRV-8833 Motor Drivers, each equipped with an Encoder. In this chapter, we first design a PID algorithm to optimize the control of the motor, and then leverage the equation of motion that was derived in Chapter 1 to build an autopilot system to optimize ALEX's locomotion. To improve efficiency, we will initially adjust the PID parameters within the Robot Operation System (ROS) environment. Subsequently, we will conduct fine-tuning under real-world operational conditions to ensure optimal system performance and adaptability.

## 2.1 Drive Platform Architecture

The DRV-8833 Motor Driver is a type of Permanent Magnet DC (PMDC) motor (see Figure 2.1(a)). As the power output of the Arduino Uno is insufficient to drive these motors, a Stepper Motor Driver Carrier Pololu A4988 must be used to channel power from the external battery pack to the motors. The Arduino is connected to the Pololu A4988, and the input power to the motors can be regulated through the Arduino via the Pulse Width Modulation (PWM) supplied to those connections.

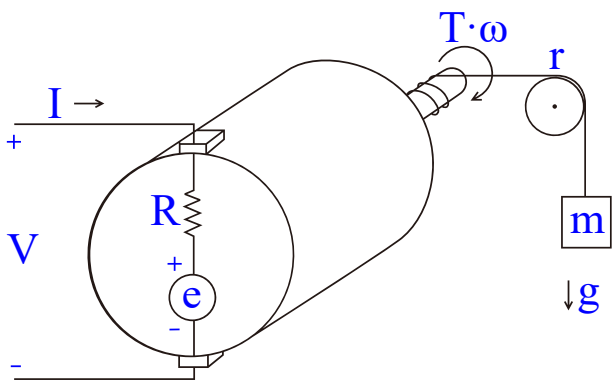
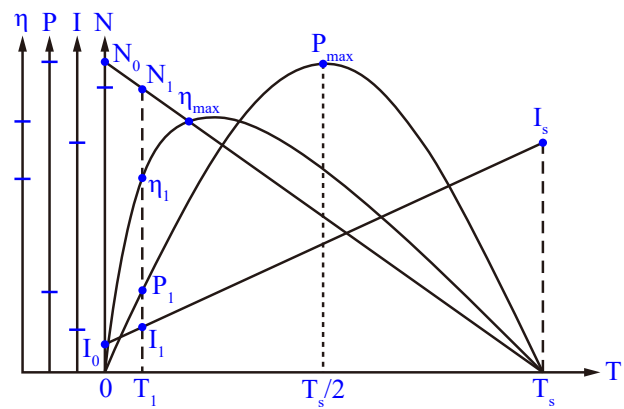

**Fig 2.1(a)** Parameters of PMDC motor
　　V (V): Terminal voltage of motor
　　I (A): Armature current
　　$\omega$ (rad/s): Rotational Speed
　　T (Nm): Torque

**Fig 2.1(b)** Characteristics Curves of motor:
　　Speed ($\omega$ or N) - Torque (T) Curve
　　Current (I) - Torque (T) Curve
　　Power output (P) - Torque (T) Curve
　　Efficiency ($\eta$) - Torque (T) Curve

Source: Jones & Flynn, Mobile Robot: Inspiration to Implementation, A K Peter, 1993.

In actual operation, we found that the motors could not drive ALEX at low torque output, indicating that the mass of ALEX is relatively heavy compared to the torque provided by the motors. Therefore, when we change the motion state of ALEX, especially when initiating motion from a stationary state, we need to ensure that the motor can provide sufficient torque to counteract the load torque. According to the motor characteristics shown in Figure 2.1(b), this requires us to increase the input power of the motor by raising the input voltage, thereby enabling the motor to generate a larger torque and avoid

stalling.[1]

When ALEX transitions from a stationary state to the desired speed, the input power is initially increased to prevent motor stalling. However, this could potentially cause ALEX to overspeed and overshoot the setpoint. This is particularly concerning because ALEX's mapping and localization, which rely on the SLAM algorithm, require the robot to move with a high degree of steadiness. Abrupt or frequent changes in motion are problematic. Therefore, it is crucial that we precisely modulate ALEX's speed. To accomplish this, we first designed a PID control algorithm for a set of motors and encoders. Then, using this initial PID control algorithm as a foundation, we designed an autopilot system to optimize the robot's motion.

## 2.2 PID Algorithm Enhanced PMDC Motor Control System

Figure 2.2 presents a block diagram of the PID controller in a feedback loop configuration. In this system, 'setEcd' represents the target degree of rotation of the motor, which is quantified in terms of encoder counts. 'rcdEcd' represents the current rotation of the motor. The difference between the target and the current rotation, denoted by 'errEcd' (or *e(t)* in Figure 2.2), is computed as the subtractive difference between 'setEcd' and 'rcdEcd'. This error value is discretely computed by the PID controller with a time difference *t′*, which then applies a correctional factor *u(t)*. This factor is a sum of the proportional (P), integral (I), and differential (D) terms as follows expressed mathematically with the following equation:

$$u(t) = k_p e(t) + k_i \int_0^{t'} e(t')\, d(t') + k_d \frac{de(t)}{dt}$$

Term proportional (P) is the instantaneous error of the system, term integral (I) is cumulative error of the system and term differential (D) is the rate of change of error of the system.

The tuning of these PID controller parameters ($k_p$, $k_i$, $k_d$) is crucial as it facilitates the desired correction, thereby achieving the target motor rotation. Finally, PWM values Vpwm are generated, enabling the Arduino to regulate the speed of the PMDC motor.

---

[1] Operated only within the Digital Systems and Applications Laboratory environment. This is because increasing the input power of the motor by raising the input voltage may lead to motor overheating or even damage.

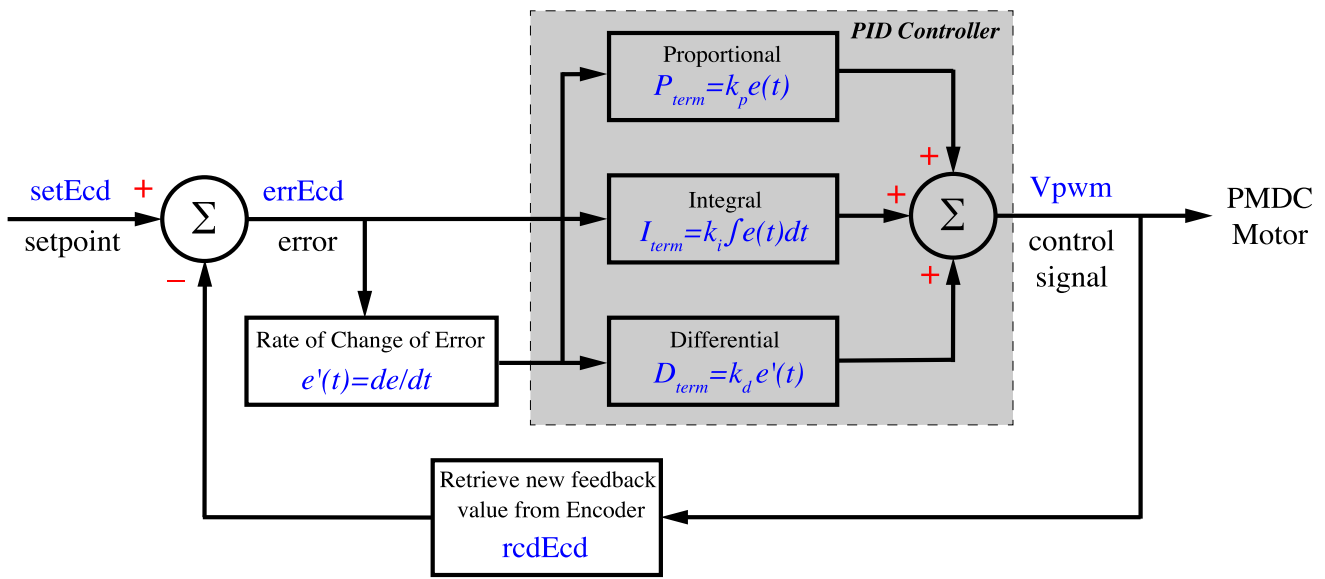

**Fig 2.2** Block diagram of a PID controller in a feedback loop

## 2.3 PID Algorithm Enhanced Locomotion

The motion equation of ALEX analyzed in Chapter 1 and the PID Algorithm Enhanced PMDC Motor Control System designed in the previous Chapter are used to fine-tune the locomotion of ALEX.

Figure 2.3 illustrates ALEX searching for victims within a maze. The red points in the figure represent the potential locations of victims and optimal stopping points (setpoints). At each point, the robot's center of rotation $O_C$ stops, and it can successfully complete turning or recognizing victim's colors. Since we have requirements for the robot's orientation during turning and color recognition, we cannot make the robot move in a circular motion as shown in Figure 1.4. Moreover, when the robot moves to the setpoint, it should stop precisely at the setpoint without overshooting. To satisfy these two conditions, the optimal motion trajectory of the robot should be as shown by the gray dashed line in Figure 2.3. The robot should first rotate in place (with the ICR as the $O_C$ point for rotation) to face the setpoint angle and then move in a straight line. Therefore, we can control the robot's turning and linear motion as two separate motion control systems – steering control system and forward control system. Decoupling these systems often leads to better stability of locomotion. It reduces the chance of instabilities that can arise from the interactions between forward and rotational motion, especially in high-speed or highly dynamic environments. For ALEX, optimal performance can be achieved by adjusting the parameters of forward and steering control system independently. For example, in a mobile robot, it might be advantageous to slow down (reduce forward speed) while turning (increasing rotational speed) to prevent tipping over.

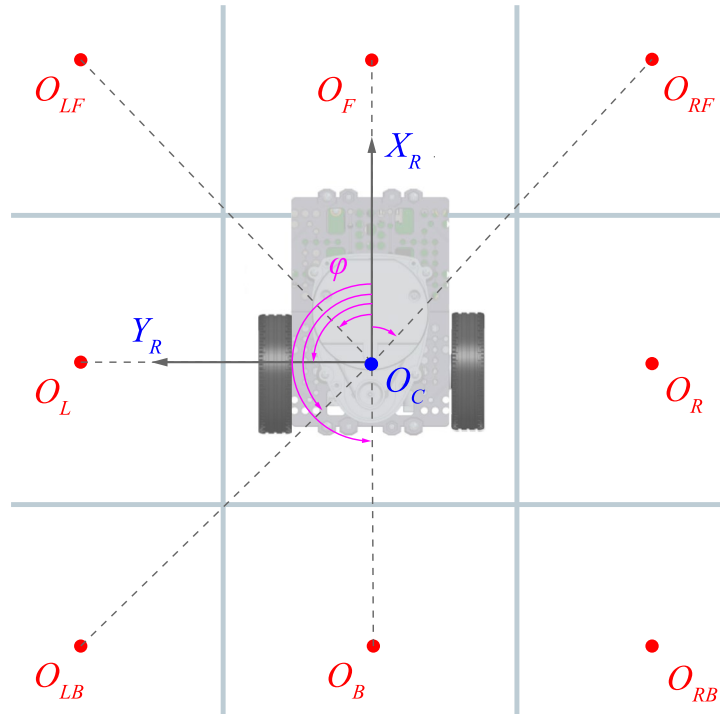

**Fig 2.3** Orientation of ALEX in rescue environment. $O_{LF}$, $O_L$, $O_{LB}$, $O_F$, $O_B$, $O_{RF}$, $O_R$, and $O_{RB}$ represent the Left Front, Left, Left Back, Front, Back, Right Front, Right, Right Back setpoint respectively. $\varphi$ is the error angle between the robot direction of motion and the setpoint.

### 2.3.1 Steering Control

When the robot revolves around the center of rotation $O_C$, its linear velocity $V_C$ should be equal to 0. Therefore, by substituting into the inverse motion equation Eq 10, we can derive the relationship between the angular velocity of rotation and the velocities of the left drive wheel $v_l$ and the right drive wheel $v_r$:

$$\begin{bmatrix} v_r \\ v_l \end{bmatrix} = \begin{bmatrix} 1 & d_w/2 \\ 1 & -d_w/2 \end{bmatrix} \begin{bmatrix} 0 \\ w \end{bmatrix} = \begin{bmatrix} w \cdot d_w/2 \\ -w \cdot d_w/2 \end{bmatrix}$$

Thus, the driving speeds of the left and right drive wheels are equal, but their directions are opposite. There exists the following relationship between the driving speed and the angular velocity:

$$v = \frac{w \cdot d_w}{2}$$

The number of rotations required for both wheels should also be equal. Therefore, after ensuring that the rotation speeds of both wheels are the same but in opposite directions, we can simply use the encoder of either the left or the right wheel to set up the PID control system.

### 2.3.2 Forward Control

When the robot moves in a straight line, the angular velocity $w$ of rotation around the center point $O_c$ should be 0. Therefore, by substituting into the inverse motion equation Eq 10, we can derive the relationship between the linear velocity of point $O_c$ and the velocities of the left drive wheel $v_l$ and the right drive wheel $v_r$:

$$\begin{bmatrix} v_r \\ v_l \end{bmatrix} = \begin{bmatrix} 1 & d_w/2 \\ 1 & -d_w/2 \end{bmatrix} \begin{bmatrix} v_c \\ 0 \end{bmatrix} = \begin{bmatrix} v_c \\ v_c \end{bmatrix}$$

The number of rotations required for both wheels should also be equal. Therefore, after ensuring that the rotation speeds and directions of both wheels are the same, we can simply use the encoder of either the left or the right wheel to set up the PID control system.

## 2.4 Autopilot System

The relative distances between the robot's center of rotation (OC) and setpoints $O_{LF}$, $O_L$, $O_{LB}$, $O_F$, $O_B$, $O_{RF}$, $O_R$, and $O_{RB}$, along with the initial angle of motion of the robot with respect to the setpoints, are pre-stored in the program's global variables in the form of encoder counts. A human operator can command ALEX to move to any setpoints, and ALEX will engage the Steering Control and Forward Control systems to move. During this motion, ALEX utilizes an Enhanced Locomotion PID Algorithm for autonomous driving.

During Autopilot mode, the robot regularly checks if the values from the left and right encoders are equal. If the discrepancy between the encoder values exceeds a defined limit, the robot disconnects from autonomous driving and is manually taken over by the human operator.

## 2.5 Tuning PID Parameters in ROS Environment

To determine the optimal parameters ($k_p$, $k_i$, $k_d$) of PID efficiently, we migrated the parameters of the real-world environment into the ROS simulation environment. Figure 2.4 (a) and (b) respectively demonstrate the debugging of the PID parameters for the Steering Control and Forward Control systems using the ROS simulation environment.

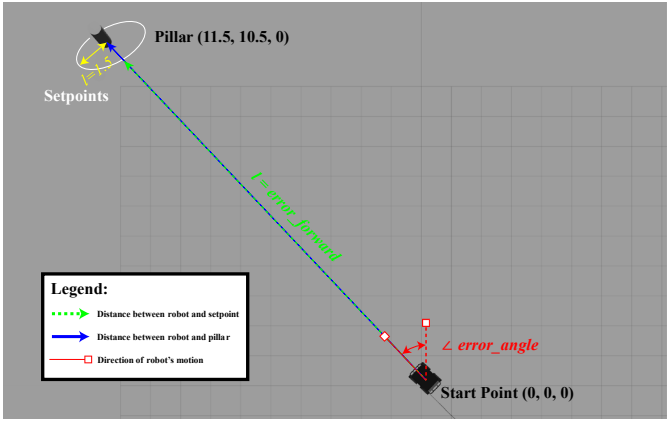 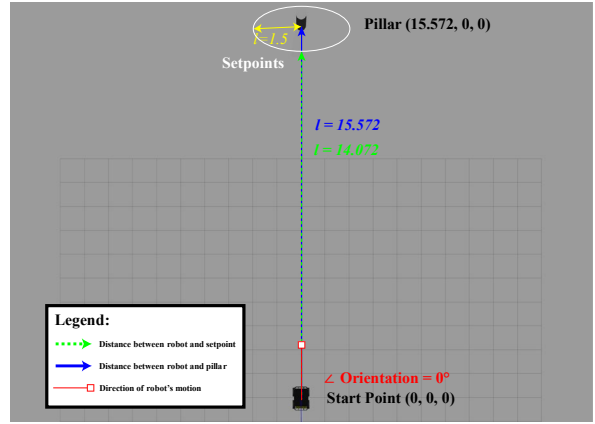

**Fig 2.4(a)** Tuning Steering Controller in ROS     **Fig 2.4(b)** Tuning Forward Controller in ROS

### 2.5.1 Tuning Methods

Tuning methods are designed to tune a PID controller by adjusting $k_p$, $k_i$ and $k_d$ parameters. In this project, PID controllers are tuned using a new method adapted from the Manual Tuning Method and the *Ziegler–Nichols* Method.

#### 2.5.1.1 Manual Tuning Method

Manual Tuning Method is based on the effects of increasing or decreasing the gain of $k_p$, $k_i$ and $k_d$ parameters separately. By analyzing the error plot curve plotted by *MatPlot*, Manual Tuning Method can be applied easily to tune parameters according to Table 2-1 below:

**Tab 2-1** Effect of increasing the gain of any one of the parameters independently.

| Parameter Increased | Rise Time | Overshoot | Settling Time | Steady-State Error | Stability |
|---|---|---|---|---|---|
| $K_p$ | Decrease | Increase | Small Change | Decrease | Degrade |
| $K_i$ | Decrease | Increase | Increase | Decrease Significantly | Degrade |
| $K_d$ | Minor Decrease | Minor Decrease | Minor Decrease | No Effect | Improve (for small $K_d$) |

The Manual Tuning Method can be relatively time-consuming because it requires increasing or decreasing each parameter independently multiple times to confirm that it is optimal. Furthermore, the tuning process is subjective and cannot be quantified. Hence, we used another heuristic tuning method called the *Ziegler–Nichols* method, introduced by *John G. Ziegler* and *Nathaniel B. Nichols* in the 1940s.

#### 2.5.1.2 Ziegler–Nichols Method

The *Ziegler–Nichols* Method is performed by setting the P, I, and D gains to zero. The P gain is then increased from zero until it reaches the ultimate gain $K_u$, at which the control loop output has stable and consistent oscillations. The oscillation will have a period of $P_u$, which will be used to set the P, I, and D gains depending on the type of controller used and the desired behavior. For this project, the gains for P, PI, and PID control systems can be set according to Table 2-2.

**Tab 2-2** Gains for P, PI, and PID control systems

| Control Type | $K_p$ | $K_i$ | $K_d$ |
| --- | --- | --- | --- |
| P | 0.50 $K_u$ | - | - |
| PI | 0.45 $K_u$ | 1.2 $K_p/P_u$ | - |
| PID | 0.60 $K_u$ | 2 $K_p/P_u$ | $K_p P_u/8$ |

**2.5.1.3 The New Tuning Method**

Although the *Ziegler–Nichols* Method is well-established, it is based on trial and error, which cannot be applied in every scenario. In this project, a new tuning method based on Manual Tuning Method and *Ziegler–Nichols* Method will be used to implement an optimized PID controller:

1. Increase $K_p$ until the forward error oscillates.
2. Half the $K_p$ approximately allows a quarter amplitude decay type response.
3. Increase $K_i$ until the offset is corrected. There is not much steady-state error to begin with.
4. Increase $K_d$ if required but not too much because it is susceptible to noise.

The forward control and steering control systems can be tuned by six parameters listed in Table 2-3 below:

**Tab 2-3** PID Parameters for Controllers

| Controller | Parameter | Description |
| --- | --- | --- |
| **Forward** | Kp_f | The proportional parameter in forward control |
|  | Ki_f | The integral parameter in forward control |
|  | Kd_f | The differential parameter in forward control |
| **Steering** | Kp_a | The proportional parameter in steering control |
|  | Ki_a | The integral parameter in steering control |
|  | Kd_a | The differential parameter in steering control |

## 2.5.2 Tuning Process

Then, we applied the new tuning method to tune the forward control system.

| Step 1: Increase $K_p$ for forward control until the forward error oscillates. ||||
|---|---|---|---|
| Forward Controller | Kp_f | 16▲ | 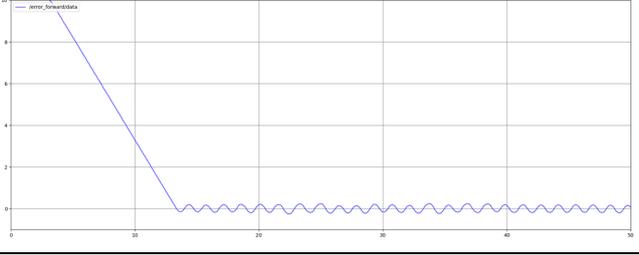 |
| | Ki_f | 0 | |
| | Kd_f | 0 | |
| Steering Controller | Kp_a | 0 | |
| | Ki_a | 0 | |
| | Kd_a | 0 | |
| Step 2: Half the $K_p$ for Forward control to allow a quarter amplitude decay response. ||||
| Forward Controller | Kp_f | 8▼ | 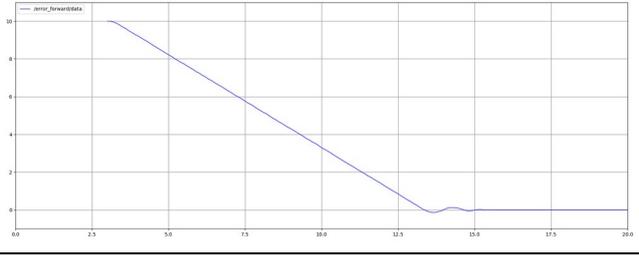 |
| | Ki_f | 0 | |
| | Kd_f | 0 | |
| Steering Controller | Kp_a | 0 | |
| | Ki_a | 0 | |
| | Kd_a | 0 | |
| Step 3: Increase $K_i$ until the offset is corrected. There is little steady-state error, to begin with. ||||
| Forward Controller | Kp_f | 8 | 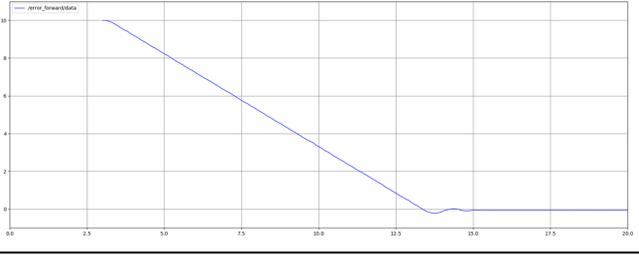 |
| | Ki_f | 0.001▲ | |
| | Kd_f | 0 | |
| Steering Controller | Kp_a | 0 | |
| | Ki_a | 0 | |
| | Kd_a | 0 | |
| Step 4: Increase $K_d$ for Forward control but not too much because it is susceptible to noise. ||||
| Forward Controller | Kp_f | 8 | 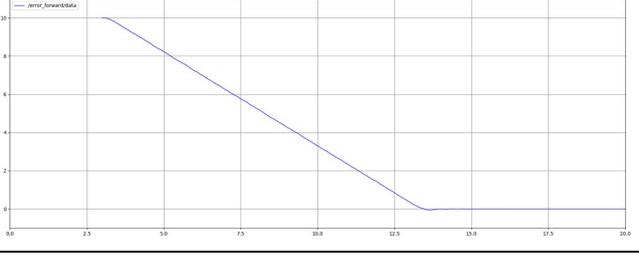 |
| | Ki_f | 0.001 | |
| | Kd_f | 1▲ | |
| Steering Controller | Kp_a | 0 | |
| | Ki_a | 0 | |
| | Kd_a | 0 | |

The forward controller's tuning procedure is completed. Then, the new tuning method is used to tune the steering control system.

| Step 1: Increase $K_p$ for steering control with the pillar at the correct *y* position until the angle error oscillates. | | | |
|---|---|---|---|
| Forward Controller | Kp_f | 0 | |
| | Ki_f | 0 | |
| | Kd_f | 0 | |
| Steering Controller | Kp_a | 40▲ | |
| | Ki_a | 0 | |
| | Kd_a | 0 | |
| Step 2: Half the $K_p$ for steering control approximately to allow a quarter amplitude decay response. | | | |
| Forward Controller | Kp_f | 0 | |
| | Ki_f | 0 | |
| | Kd_f | 0 | |
| Steering Controller | Kp_a | 20▼ | |
| | Ki_a | 0 | |
| | Kd_a | 0 | |
| Step 3: Increase $K_d$ for steering control but not too much because it is susceptible to noise. | | | |
| Forward Controller | Kp_f | 0 | |
| | Ki_f | 0 | |
| | Kd_f | 0 | |
| Steering Controller | Kp_a | 20 | |
| | Ki_a | 0 | |
| | Kd_a | 1.35▲ | |

Applying the PID parameters obtained in the ROS simulation environment to the real-world robot control has resulted in quicker response times when the robot reaches the setpoint, smoother velocity changes, and prevention of overshooting the designated point.

However, in real-world rescue environments, the robot often encounters uneven maze surfaces, leading to a large discrepancy between the left and right encoder readings, which in turn disconnects autonomous driving. To enhance the robot's obstacle-crossing ability and reduce the frequency of disconnections, a PID controller optimized using a fuzzy algorithm will be designed and applied to improve ALEX robot's movements, which will be covered in Chapter 3.

# Chapter 3 Fuzzification of Autopilot System

Uneven terrain is a significant factor that leads to the disconnection of a robot's autopilot system. Small hills within the maze may cause the robot to deviate from its set motion direction during movement or cause the robot's drive system to lack enough torque to drive the robot over the hills. When the robot encounters the above situations, it often triggers the conditions for disconnection of the autopilot designed in Chapter 2.4. This chapter will fuzzify the Autopilot system designed in Chapter 2, using a fuzzy PID control system to overcome obstacles and avoid the need for manual operations during obstacle crossing.

Fuzzification of the Autopilot System is the use of a fuzzy controller for online self-tuning of the parameters of the PID controller. The process is as follows: first, find out the fuzzy relationship between the three PID parameters $k_p$, $k_i$, $k_d$, and the error $E$ and rate of change of error $EC$. Then, by continuously detecting the error and the change in error during operation, adjust the three parameters online according to the principles of fuzzy control, to meet the different requirements for controller parameters at different errors and error rate changes. Figure 3.1 presents a block diagram of the Fuzzy - PID controller in a feedback loop configuration:

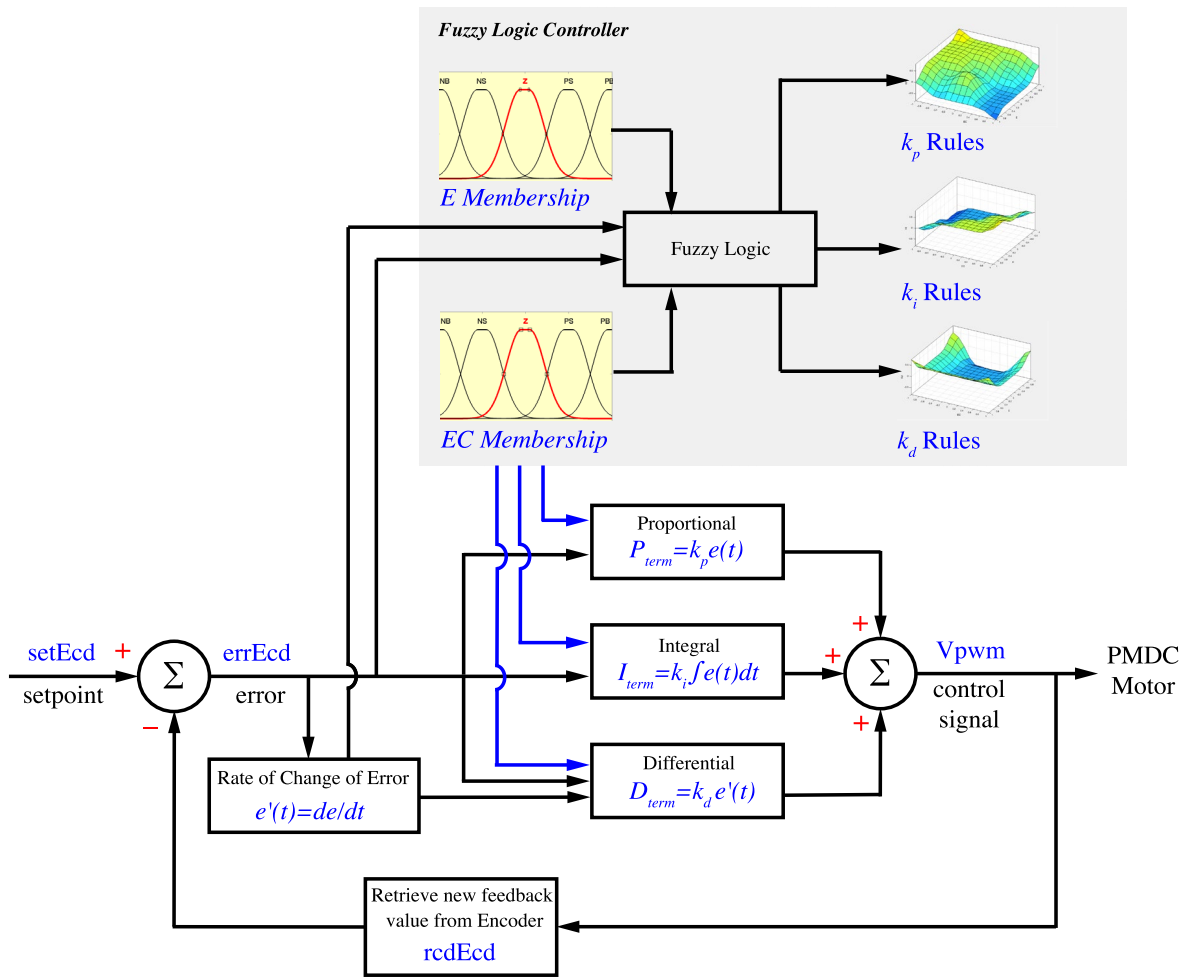

**Fig 3.1** Block diagram of a Fuzzy - PID controller in a feedback loop

To find out the fuzzy relationship between the three PID parameters and the error and rate of change of error, we first define the membership function of them.

## 3.1 Membership Functions
### 3.1.1 Input Functions
This paper defines seven fuzzy sets on various parameter fuzzy domains: Negative Big (NB), Negative Medium (NM), Negative Small (NS), Zero (ZO), Positive Small (PS), Positive Medium (PM), and Positive Big (PB)

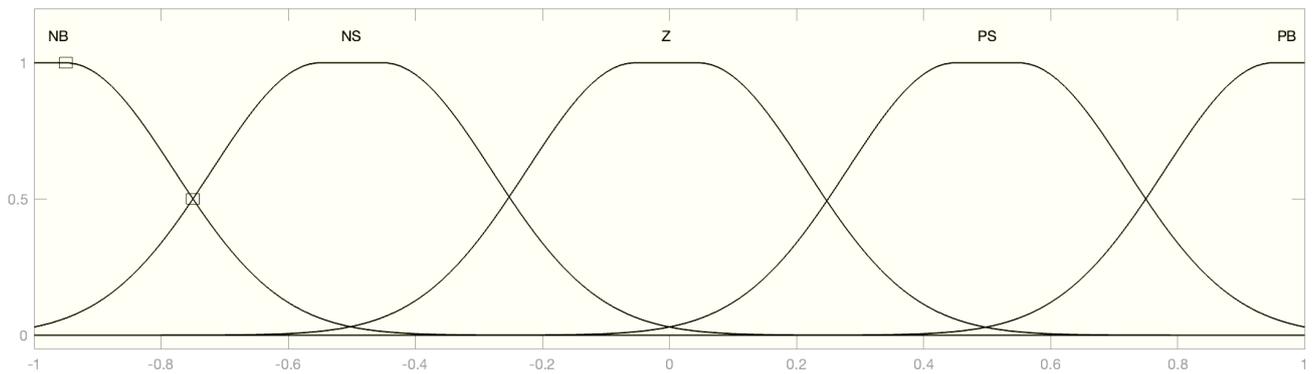

**Fig 3.2** Error Membership Functions

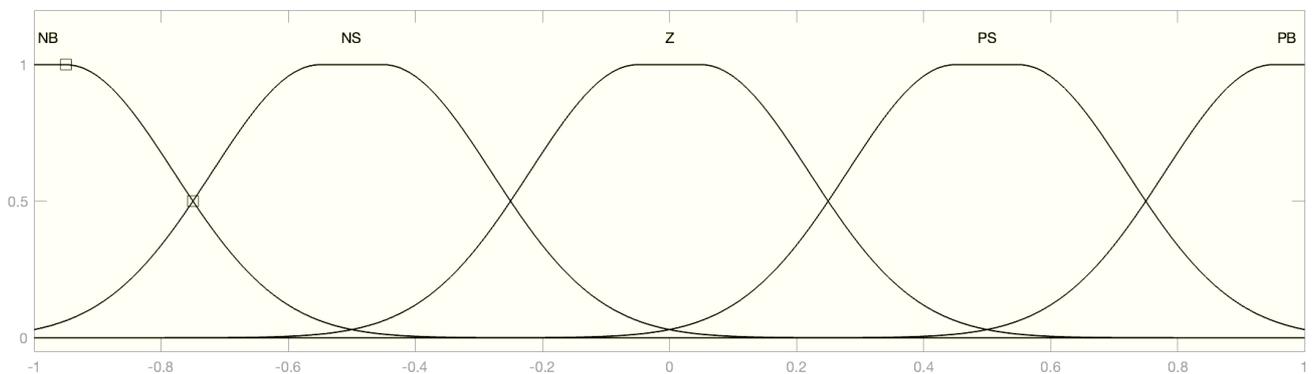

**Fig 3.3** Rate of Change of Error Membership Functions

### 3.1.2 Output Functions

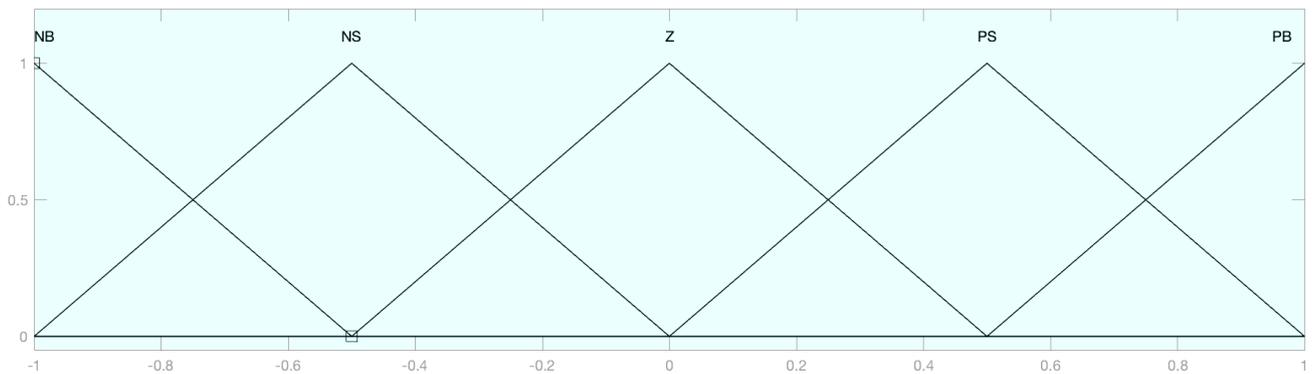

**Fig 3.4** $K_p$ Membership Functions

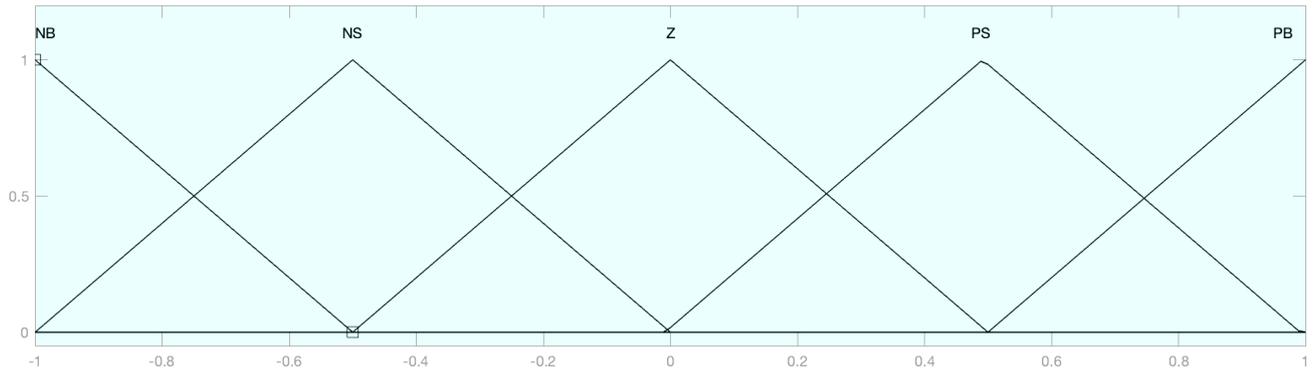

**Fig 3.5** $K_i$ Membership Functions

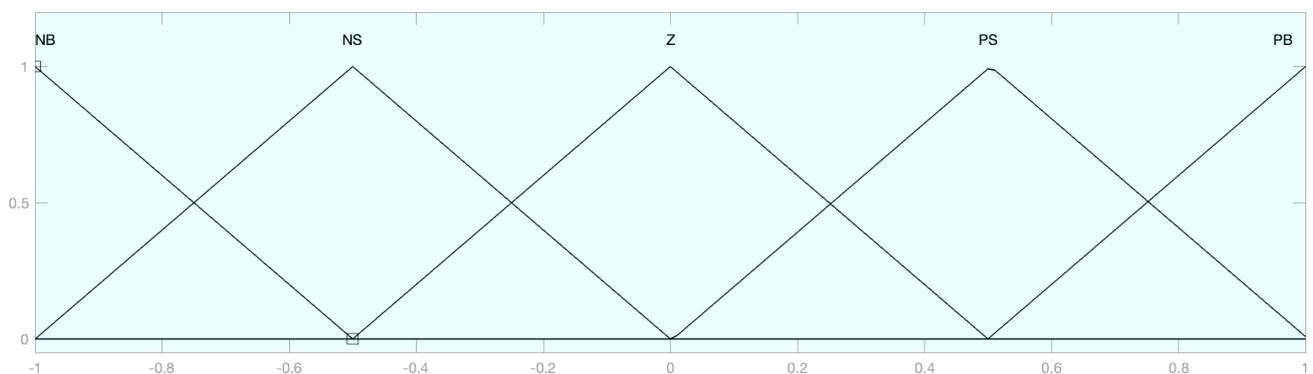

**Fig 3.6** $K_d$ Membership Functions

## 3.2 Implementation of Fuzzy Rules

When the robot is climbing a slope, the drive motor needs to provide more torque to successfully overcome the hill because the robot needs to overcome more gravity. Because kinetic energy is partially converted into gravitational potential energy, the robot's speed will slow down when going uphill.

When the robot is coming down from a slope, the robot's gravitational potential energy is converted into kinetic energy, which can cause the robot to exceed its speed limit. To control the speed and prevent overshooting, the drive motor needs to decelerate.

Based on the above rules, we have formulated the following Fuzzy Rule to generate better PID parameters to control robot movement.

**Tab 3.1** $E – EC – K_p$ Rules

| E \ EC | NB | NM | NS | ZO | PS | PM | PB |
|---|---|---|---|---|---|---|---|
| NB | PB | PB | PM | PM | PS | ZO | ZO |
| NM | PB | PB | PM | PS | PS | ZO | NS |
| NS | PM | PM | PM | PS | ZO | NS | NS |
| ZO | PM | PM | PS | ZO | NS | NM | NM |
| PS | PS | PS | ZO | NS | NS | NM | NM |
| PM | PS | ZO | NS | NM | NM | NM | NB |
| PB | ZO | ZO | NM | NM | NM | NB | NB |

**Fig 3.7** Visualization of $E – EC – K_p$ Rules

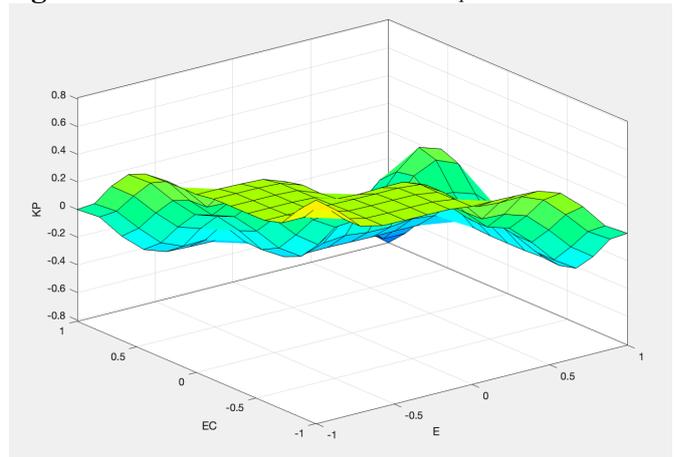

**Tab 3.2** $E – EC – K_i$ Rules

| E \ EC | NB | NM | NS | ZO | PS | PM | PB |
|---|---|---|---|---|---|---|---|
| NB | NB | NB | NB | NB | NM | ZO | ZO |
| NM | NB | NB | NB | NB | NM | ZO | ZO |
| NS | NM | NM | NM | NM | ZO | PS | PS |
| ZO | NM | NM | NS | ZO | PS | PM | PM |
| PS | NS | NS | ZO | PM | PM | PM | PM |
| PM | ZO | ZO | PM | PB | PB | PB | PB |
| PB | ZO | ZO | PM | PB | PB | PB | PB |

**Fig 3.8** Visualization of $E – EC – K_i$ Rules

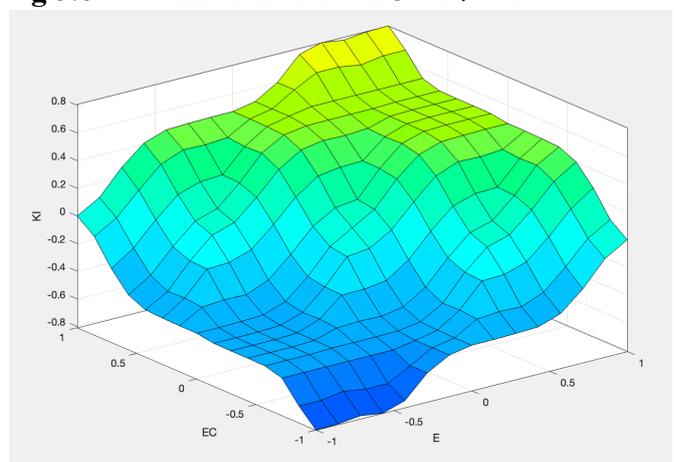

**Tab 3.3** $E – EC – K_d$ Rules

| E \ EC | NB | NM | NS | ZO | PS | PM | PB |
|---|---|---|---|---|---|---|---|
| NB | PS | NS | NB | NB | NB | NM | PS |
| NM | PS | NS | NB | NM | NM | NS | ZO |
| NS | ZO | NS | NM | NM | NS | NS | ZO |
| ZO | ZO | NS | NS | NS | NS | NS | ZO |
| PS | ZO | ZO | ZO | ZO | ZO | ZO | ZO |
| PM | PB | NS | PS | PS | PS | PS | PB |
| PB | PB | PM | PM | PM | PS | PS | PB |

**Fig 3.9** Visualization of $E – EC – K_d$ Rules

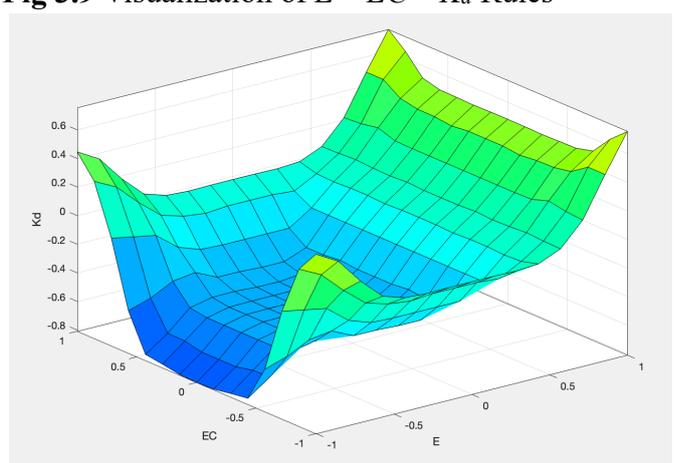

## 3.3 Defuzzification of Fuzzy Controllers in Simulink®

This paper uses the Fuzzy logic editor based on MATLAB Simulink to defuzzify the designed Fuzzy Controllers. The following image is a screenshot of the Simulink software. At this moment, both the error and the rate of change of error are 0. After defuzzification, the PID parameters generated by the Fuzzy Controller are also close to 0.

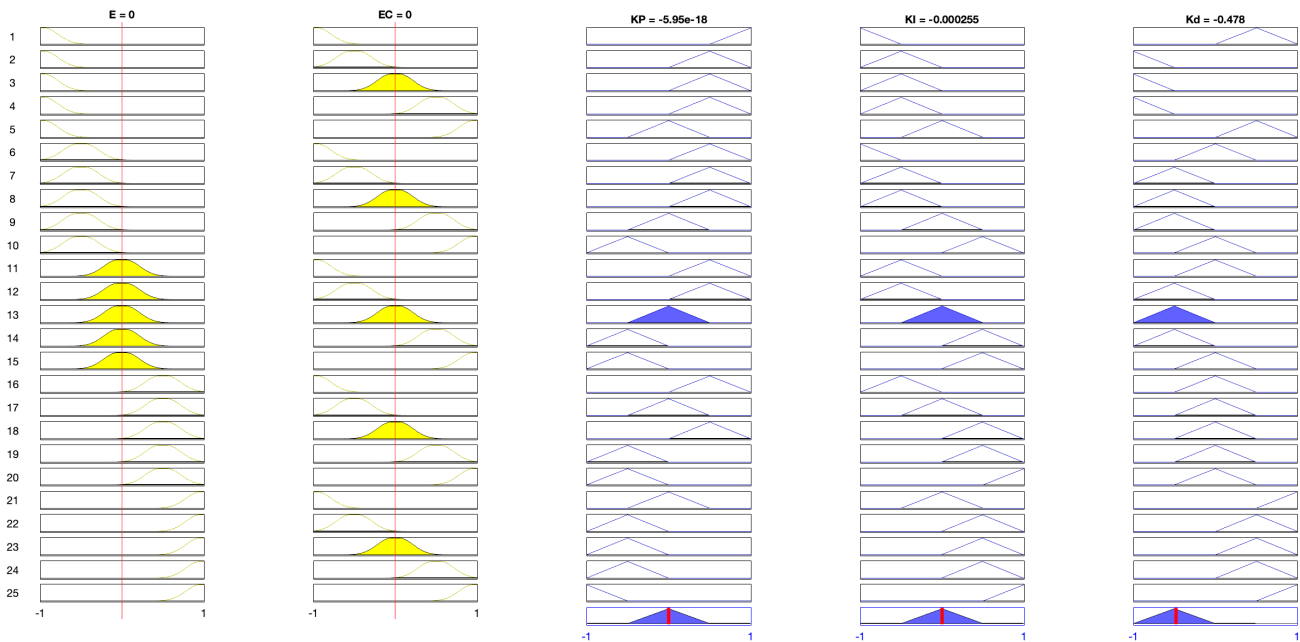

# Chapter 4 Fuzzification of Color Detection Algorithm

There are inherent difficulties with the color sensor that cannot be completely controlled. Factors such as lighting and shadows, and even the quantum electrical effects in the sensor chip, make it virtually impossible to ensure that the color being analyzed remains constant when the robot is detecting the color of the victim. Since color is greatly influenced by many potential factors, fuzziness is incorporated into the system to address the uncertainty problem of color object classification.

Given these inherent challenges in color detection—factors like lighting, shadows, and quantum electrical effects causing uncertainty—fuzziness, or fuzzy logic, proves to be a valuable asset. Fuzzy logic mimics human decision-making by considering degrees of truth, rather than absolute binary values. This allows the color detection system to express uncertainty, effectively stating that a color is, for instance, "75% similar to red" instead of being strictly "red" or "not red". This is accomplished by defining fuzzy sets for each color, which include possible color variations and their degree of membership to the original color. When a color is detected, the system determines its membership to these sets. To translate these fuzzy values back to a singular, definitive color decision—a process known as defuzzification—the system typically selects the color with the highest membership value. Therefore, the incorporation of fuzziness into the system is a logical way to address the uncertainty problem in color identification in robotics.

## 4.1 Color Detector Architecture

The TCS3200, a color sensor capable of detecting and measuring the intensity of red, green, and blue (RGB) light, operates via a grid of photodiodes with color filters, converting light into an electrical signal relative to light intensity. This sensor, housing an 8×8 array of photodiodes with 16 of each type for red, green, blue, and clear (no filter) filters, uses an internal oscillator to generate a frequency signal that is output as a square wave on the sensor's output (OUT) pin, with the frequency of the output signal proportional to the detected light intensity. When integrated with the Raspberry Pi via its onboard General Purpose Input Output (GPIO) pins - with the wiring architecture demonstrated in Figure 4.1 below - the TCS3200's output frequency corresponding to each color's intensity can be measured by connecting the OUT pin to a Raspberry Pi GPIO pin, recording the time between the square wave's rising and falling edges, thereby obtaining the intensity values for color recognition.

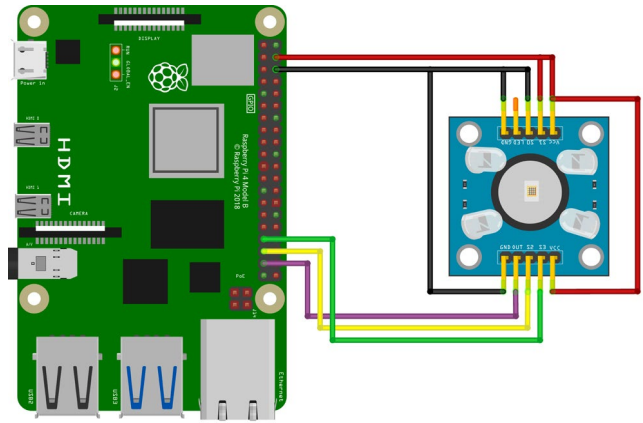

**Fig 4.1** Wiring Architecture Chart

Through the above properties of the TCS3200 color sensor, we can run a Python script on Raspberry Pi to obtain the reflected light intensity of RGB color. Here is a pseudocode to show the flowchart of color detection:

| Pseudocode |
| --- |
| ```
1. START
|
2. Import RPi.GPIO as GPIO, import time
|
3. Define pins (s2, s3, out), NUM_CYCLES
|
4. Define setup() function
|    └── Set GPIO mode
|    └── Set up GPIO pins
|
5. Define read_value(a0, a1) function
|    └── Set GPIO outputs for s2 and s3
|    └── Sleep 0.1 seconds
|    └── Wait for GPIO falling edge, then rising edge
|    └── Start time measurement
|    └── Wait for GPIO falling edge
|    └── Calculate and return elapsed time
|
6. Define loop() function
|    └── Infinite loop
|        ├── Print red value
|        ├── Sleep 0.1 seconds
|        ├── Print green value
|        ├── Sleep 0.1 seconds
``` |





(continued)

```
|           ├── Print blue value
|           └── Sleep 1 second
|
7. MAIN
|   ├── Call setup() function
|   ├── Try loop()
|   └── Handle KeyboardInterrupt, cleanup GPIO
|
8. END
```

For the light intensity of each color returned by the sensor, we can develop a color recognition algorithm based on fuzzy logic.

## 4.2 Fuzzified Color Detection

Through continuous calibration and adjustment of color data in the Digital Systems and Applications laboratory, we obtained the following comparison table of color and light intensity value $x_{R/G/B}$ at specific detection distances (from 2cm to 8cm). Additionally, we have calculated the RGB values and fuzzy logic membership degrees for each color.

**Tab 4.1** Comparison of color and intensity

| Color | Far (8 cm) | | | Near (2 cm) | | |
|---|---|---|---|---|---|---|
| | **R** | **G** | **B** | **R** | **G** | **B** |
| **Black** | 571 | 527 | 364 | 433 | 390 | 263 |
| **Green** | 355 | 296 | 258 | 199 | 125 | 138 |
| **Red** | 252 | 399 | 258 | 92 | 234 | 170 |
| **Orange** | 197 | 302 | 253 | 80 | 166 | 151 |
| **Blue** | 367 | 286 | 191 | 186 | 94 | 52 |
| **Purple** | 294 | 306 | 202 | 128 | 112 | 60 |
| **White** | 85 | 80 | 59 | 51 | 41 | 34 |

In Chapter 4.2.1, we will set up membership functions between light intensity and color. The horizontal axis represents light intensity under red, green, or blue filters, while the vertical axis represents membership degrees. The membership function of each color is represented by a triangle, in which the height of the triangle indicates the membership degree. The two points at the base of the triangle represent the minimum and maximum intensity thresholds for the color, while the point at the top signifies the RGB value at which the color intensity reaches the highest membership degree. The fill color of each triangle represents the color of the corresponding fuzzy membership function.



Chapter 4 Fuzzification of Color Detection Algorithm                                            Page 27 / 35### 4.2.1 Membership Functions

The triangular membership function is used in this project because it is a common function form in fuzzy logic and can well represent the transition between different membership degrees. In this case, the membership function we use is based on experience and experimentation, but in practical applications, the choice of membership function may vary depending on the specific application.

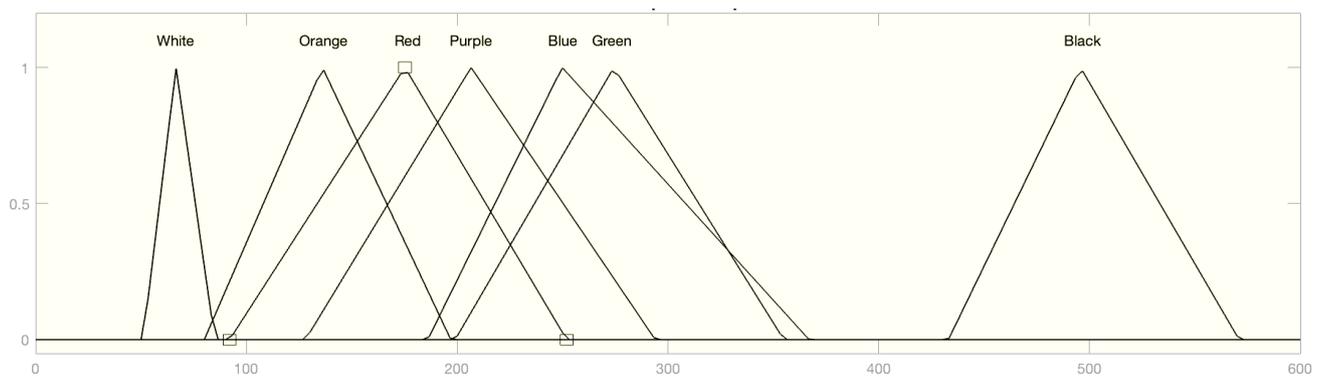

**Fig 4.2** Light intensity (under red filter) and color membership function

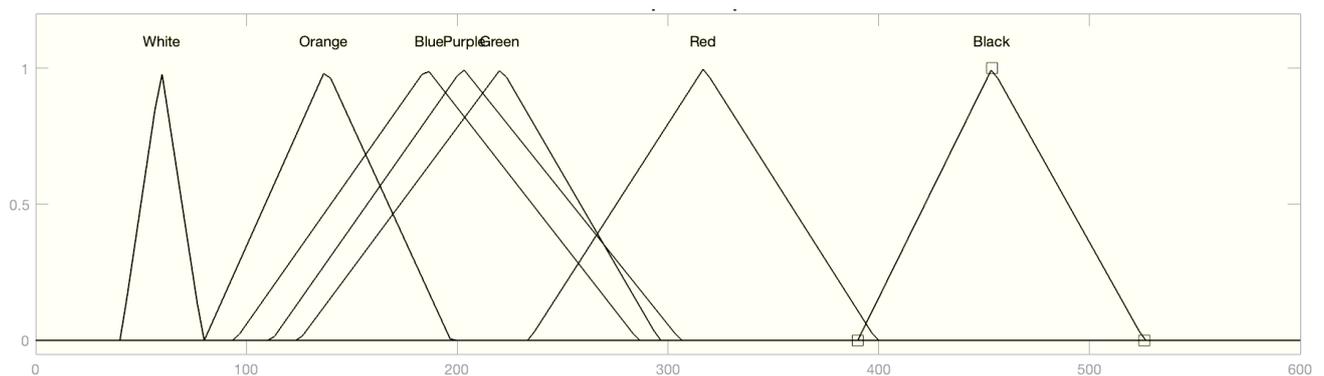

**Fig 4.3** Light intensity (under green filter) and color membership function

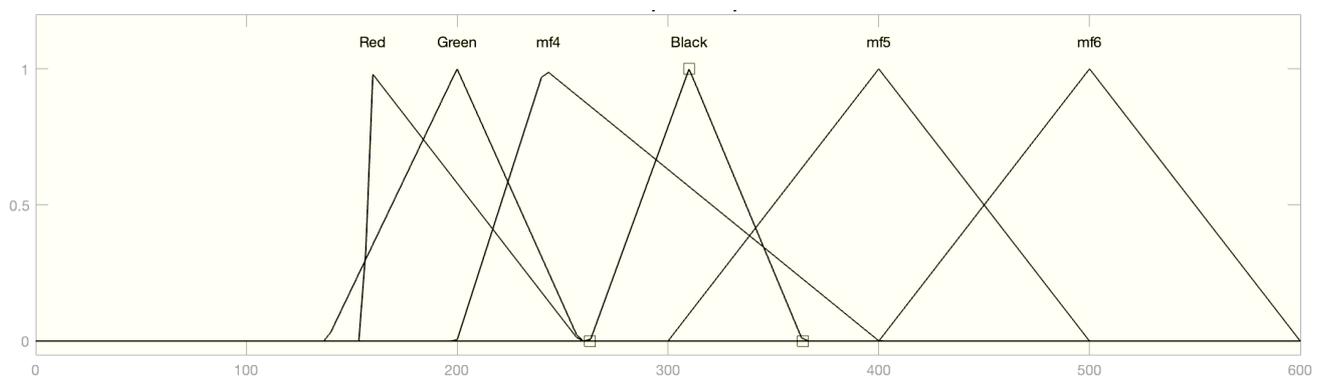

**Fig 4.4** Light intensity (under blue filter) and color membership function

guoyi@ieee.org                          First Year Project Report - AY2022/23                          Chen Guoyi



**4.2.2 Defuzzification of Fuzzy Color Detection**

Below is an example of a fuzzy rule set based on fuzzy logic membership degrees (see Table 4.2), used to determine the color category to which an RGB color value belongs. The fuzzy rule set has a total of 7 color categories, and the fuzzy rule for each category is determined by the membership degrees of the R, G, and B color components. For example, in the fuzzy rule for black, the membership degrees of R, G, and B are "Low, Low, Low," indicating that they belong to the "low" membership degree of the three-color components, so it is judged to be black. The values of each membership degree can be modified according to the actual situation.

**Tab 4.2** Fuzzy rule set for color recognition.

|        | R   |     |      | G   |     |      | B   |     |      |
|--------|-----|-----|------|-----|-----|------|-----|-----|------|
|        | LOW | MED | HIGH | LOW | MED | HIGH | LOW | MED | HIGH |
| Black  | 1   | 0   | 0    | 1   | 0   | 0    | 1   | 0   | 0    |
| Green  | 0   | 1   | 0    | 0   | 1   | 0    | 0   | 1   | 0    |
| Red    | 1   | 0   | 0    | 0   | 1   | 0    | 1   | 0   | 0    |
| Orange | 1   | 0   | 0    | 0   | 1   | 0    | 0   | 1   | 0    |
| Blue   | 0   | 1   | 0    | 1   | 0   | 0    | 1   | 0   | 0    |
| Purple | 0   | 1   | 0    | 1   | 0   | 0    | 0   | 1   | 0    |
| White  | 0   | 1   | 0    | 0   | 1   | 0    | 0   | 1   | 0    |

Based on the information provided in the table above, we can derive the following predicate logic for determining colors:

```
IF  R is Low   AND  G is Low   AND  B is Low   THEN  the color is Black
IF  R is Med   AND  G is Med   AND  B is Med   THEN  the color is Green
IF  R is Low   AND  G is Med   AND  B is Low   THEN  the color is Red
IF  R is Low   AND  G is Med   AND  B is Med   THEN  the color is Orange
IF  R is Med   AND  G is Low   AND  B is Low   THEN  the color is Blue
IF  R is Med   AND  G is Low   AND  B is Med   THEN  the color is Purple
IF  R is Med   AND  G is Med   AND  B is Med   THEN  the color is White
```





## 4.3 Limitation of Fuzzified Color Detection

As we consider Table 4.2 as a 7×9 two-dimensional fuzzy logic matrix, we find that the rank of this matrix is 6, which is not equal to the number of colors (7) that the sensor needs to recognize. Upon inspection, we notice that the predicate logic conditions for determining red and white colors are the same, i.e., the R, G, and B membership degrees for green and white in the fuzzy rules are "Medium, Medium, Medium." If we judge the colors based on the aforementioned fuzzy rules, it is possible to confuse red with white. Fortunately, during ALEX robot's rescue operation, it will not encounter white dummies, so we can use the fuzzy rules described above to determine colors in a laboratory environment. Due to the difficulty in fixing the detection distance when robots are sensing colors of victims, the R, G, and B values detected by the TCS3200 sensor may undergo slight changes, which could potentially lead to erroneous color detection results. By applying fuzzy logic to the system, strict color definitions are relaxed, and flexibility of detecting distance (2cm – 8cm) is incorporated into the system by defining colors based on fuzzy sets and rules.





# Appendix: Hardware Architecture

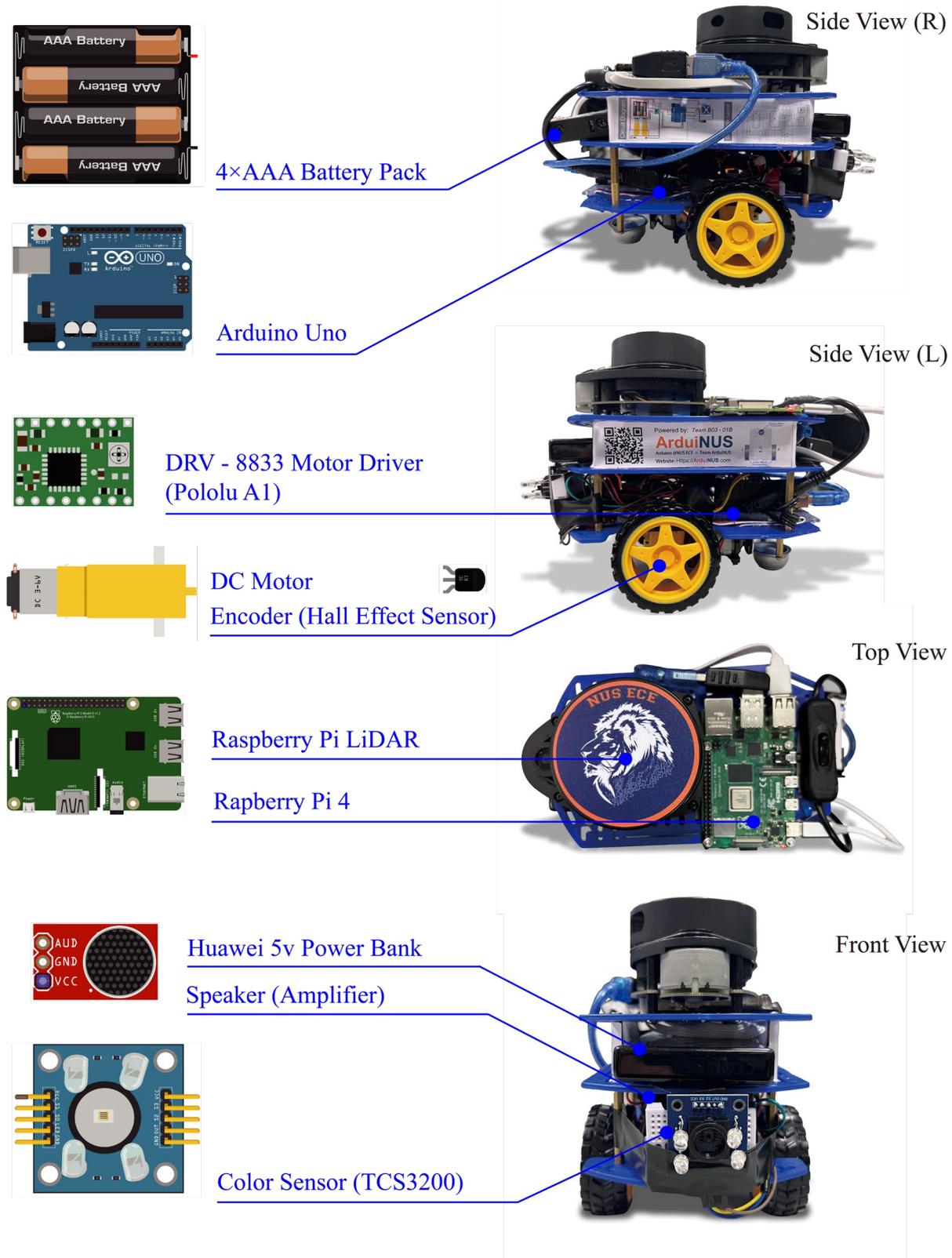

**Fig I.1** Three View Chart





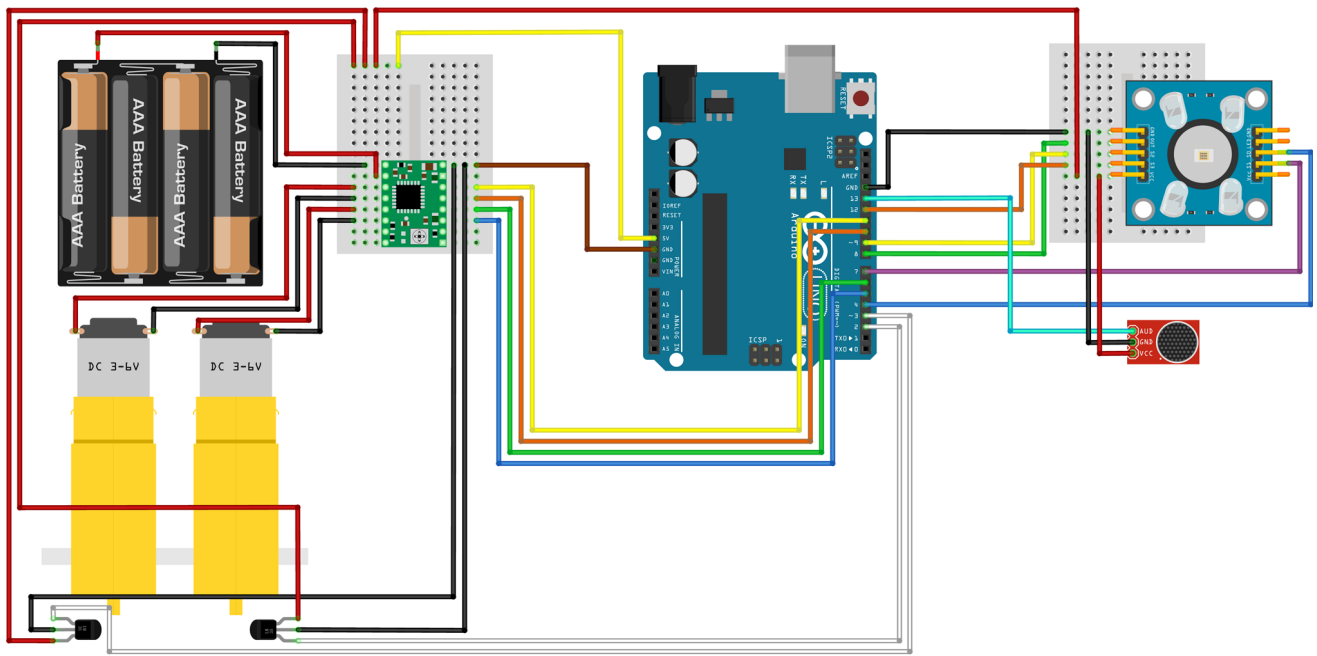

**Fig I.2** Wiring Architecture Chart

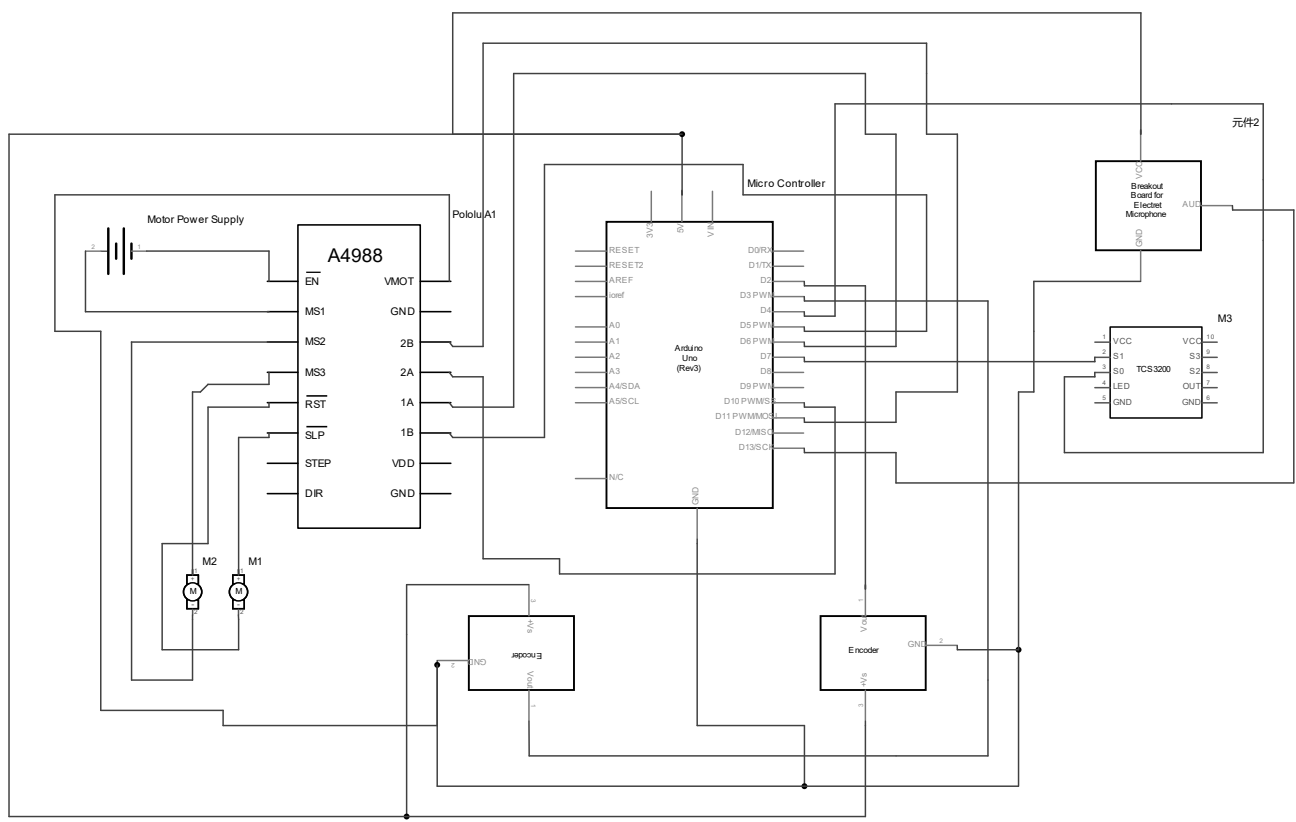

**Fig I.3** PIN/PORT Mapping Chart





# Acknowledgment

*"What kind of robots will you design?"*

Prof. Chang asked me this question in the interview for the Science & Technology Scholarship. With a strong desire to figure out the answer to the question, I came to my first semester at NUS, enrolled in module EE3305/ME3243 at the suggestion of Mr. Shi Zhansen, a warm-hearted senior. This decision became the foundation for this First Year Project (FYP) report. I titled this acknowledgement "*Guoyi's First Adventure in Robotics*," a testament to my invaluable experiences throughout my research endeavor, nestled at the end of this report.

The senior recommended I learn module EE3305/ME3243 because he considered that 'improvement should be your objective as you have a basic command of robotic knowledge.' Before being a student of NUS, I participated in *RoboCup* on behalf of my home country and have already applied the PID control algorithm in optimizing robot's locomotion. But the PID tuning strategy used previously failed to tune the parameters systematically and strategically; rather, it was based entirely on experience and guessing. The lectures delivered by Prof. Vadakkepat and Dr. Putra during module EE3305 introduced me to a more thoughtful tuning strategy which I tested intensively with the ROS system. My academic report submitted for this module, *Tuning Strategy of PID Parameters for Coupled PID Control Systems and its Applications*, is the summary of the strategy.

*"The curriculum (Computer Engineering) aims to bring real-world problems, solutions, and experiences into the university environment. "*

This is a quote from NUS official website of my home course, Computer Engineering, which is also the conclusion of my research process in my first semester. To be specific, I pondered the mutual interference between the outputs of the coupled control systems in ROS environment as a practical problem rather than ignoring it. And support from lecturers and teaching assistants was well received in the research process. In particular, my discussions with Dr. Guo Haoren and Dr. Xian Yuanjie from the NUS Control & Simulation Laboratory helped me learn about the decoupling method by analyzing motion equations. On this basis, a tuning strategy for the coupled PID control systems was developed in combination with my driving experience in real life.

After completing my first semester, I continued my exploration in the field of robotics. I had the privilege to accompany Dr. Zhang Yang at the 2023 Computer Science Research Week held at NUS School of Computing. This event provided me with the opportunity to engage with renowned computer scientists around the world. One scholar in the field of robot vision remarked that my solution during EE3305 bore similarities to fuzzy neural theory. Encouraged and assisted by these scholars, I began exploring fuzzy neural theory. As a freshman, I could only audit the EE4305 module lectures and independently sought related materials on fuzzy theory in the NUS library.





> *"…I am extremely enthusiastic to become an NUS student and a life-long learner under the Interdisciplinary Learning Scheme…"*

My experience of self-studying fuzzy neural theory and my first experience of independently reading literature made me understand how important the lectures and experiments carefully prepared by the professors are. Thanks to the Lecture Notes made by Prof. Xiang for module EE4305, the content he summarized significantly improved my efficiency in learning fuzzy control systems. After studying the content of engineering linear algebra in the second semester, I reviewed the lectures of Prof. Chew Chee Meng in module ME3243. I successfully derived the motion equation of the ALEX robot drive system according to the examples he provided, thus using a more rigorous mathematical method to decouple the robot drive control system. These laid a solid foundation for me to reproduce the fuzzy control system using the ALEX robot in the real environment of module CG2111A.

> *"My aim is to bring the knowledge I learned in university to solve real-world problems."*

Like the Fuzzy - PID control system, my aim of learning and the teaching objectives of my home course also turn into a closed loop. I have solved real-world problems by applying the knowledge and experience I gained from learning and thinking in my campus life. This FYP report was cited by the *RoboCup* national team developing *A Robust PID Line Patrol Algorithm Based on Four Optical Sensors*. The successful completion of this paper would not have been possible without the help of Ms. Arcinas Camille Esther Walet, an academic writing tutor from Yale-NUS. She accompanied me through mock presentation, correcting and improving the grammar and logical errors in my paper writing. I am grateful for her suggestions for the wording of this paper and for her attitude in reviewing the paper from her perspective of different professional backgrounds.

My experience in taking modules EE3305, ME3243, EE4305, and CG2111A comprise my First Adventure in Robotics, which keeps my resolve in the discipline of Robotics during exploration. I was fortunate to get a preliminary understanding of the subject I loved in my fresh year and began to learn Core Modules according to my plan. I am looking forward to joining the Undergraduate Research Opportunities Programme and continuing to indulge my passion for the research I am interested in after certain knowledge accumulation.

In the end, to answer the questions Prof. Chang asked me when I graduated from high school, I realize I have a long journey ahead. However, I am confident that three years of knowledge accumulation at NUS will provide a satisfactory answer.

> 路漫漫其修远兮，吾将上下而求索。
> The way ahead is long; I see no ending, yet high and low I will search with my will unbending.

Chen Guoyi
15 June 2023
At the National University of Singapore